\documentclass[conference]{IEEEtran}
\usepackage{flushend}

\usepackage{url}
\usepackage{array}
\usepackage{graphicx}
\usepackage[caption=false]{subfig}
\usepackage{amsmath}
\usepackage{mdwlist}
\usepackage{multirow} 
\usepackage{algorithm,algpseudocode}
\usepackage{setspace}
\let\Algorithm\algorithm
\renewcommand\algorithm[1][]{\Algorithm[#1]\setstretch{1.2}}
\DeclareMathOperator{\arccot}{arccot}

\begin{document}
\title{On a Feedback Control-based Mechanism of Bidding for Cloud Spot Service}


\author{\IEEEauthorblockN{Zheng Li\IEEEauthorrefmark{1},
Maria Kihl\IEEEauthorrefmark{1} and
Anders Robertsson\IEEEauthorrefmark{2}}
\IEEEauthorblockA{\IEEEauthorrefmark{1}Department of Electrical and Information Technology\\
Email: \{Zheng.Li, Maria.Kihl\}@eit.lth.se}
\IEEEauthorblockA{\IEEEauthorrefmark{2}Department of Automatic Control\\
Email: Anders.Robertsson@control.lth.se\\
Faculty of Engineering, Lund University, Lund, Sweden}}

\maketitle

\begin{abstract}
As a cost-effective option for Cloud consumers, spot service has been considered to be a significant supplement for building a full-fledged market economy for the Cloud ecosystem. However, unlike the static and straightforward way of trading on-demand and reserved Cloud services, the market-driven regulations of employing spot service could be too complicated for Cloud consumers to comprehensively understand. In particular, it would be both difficult and tedious for potential consumers to determine suitable bids from time to time. To reduce the complexity in applying spot resources, we propose to use a feedback control to help make bidding decisions. Based on an arccotangent-function-type system model, our novel bidding mechanism imitates fuzzy and intuitive human activities to refine and issue new bids according to previous errors. The validation is conducted by using Amazon's historical spot price trace to perform a set of simulations and comparisons. The result shows that the feedback control-based mechanism obtains a better trade-off between bidding rationality and success rate than the other five comparable strategies. Although this mechanism is only for black-box bidding (price prediction) at this current stage, it can be conveniently and gradually upgraded to take into account external constraints in the future.
\end{abstract}


\IEEEpeerreviewmaketitle

\section{Introduction}
\label{sec:introduction}
As a key role in the success of Cloud computing in industry \cite{Weinhardt_Anandasivam_2009}, various pricing techniques have been employed by Cloud providers to attract consumers and sell Cloud services. In the de facto Cloud market, the existing pricing schemes can be generally distinguished between fixed pricing and spot pricing. Particularly, the fixed pricing schemes for both on-demand services and reserved services are dominant approaches to trading Cloud resources nowadays \cite{Al-Roomi_2013,Xu_Li_2013}. However, given the normally unpredictable
and stochastic demand, there would always
be unused resources in the virtually infinite compute capacity
of the Cloud. To further help and better utilize the idle compute
resources, spot pricing has been realized as a promising scheme that could attract more demands with toleration
of service delay and interruptions \cite{Wang_Qi_2013}, as launched by Amazon in December 2009 \cite{Song_Yao_2013}.

Unlike the static and straightforward prices driven by the fixed pricing schemes, the spot price varies based on the supply and demand of available capacity of a Cloud spot service. %In such a way, Cloud resources can be used more efficiently and even generate extra revenue for Cloud providers.
For Cloud consumers, such a spot pricing scheme would deliver various benefits ranging from economic advantages for flexible workloads to accelerations of small-scale jobs. For example, the quantitative analyses of Amazon's
price trace show that consumers can expect to save more than
half the expense if replacing on-demand instances with the spot
ones  \cite{Ben-Yehuda_2013,Javadi_Thulasiram_2013,Song_Yao_2013}. The empirical studies deliver
even more encouraging results: with proper bids, the total
cost of employing spot resources can be maintained between
13\% and 36\% of using the equivalent on-demand resources \cite{Leslie_Lee_2013,Mattess_Vecchiola_2010}. Moreover, by employing
additional spot nodes in the MapReduce process, the speedup for the overall MapReduce
time of some workloads can exceed 200\% with an extra monetary
cost of 42\% only \cite{Chohan_Castillo_2010}.

Nevertheless, it seems that potential Cloud consumers are still hesitating to enter the Cloud spot market. To employ a spot service, Cloud consumers need to submit bids on particular resources. The resource requests can be granted when the corresponding bid exceeds the current spot price, while the employed service will be interrupted when the spot price exceeds the current bid. As such, the potential consumers have to make trade-off decisions between availability and cost when using the spot service \cite{Ostermann_Prodan_2012}, which could be both challenging and tedious. For instance, by observing historical spot prices, a natural thought could be using high-enough bids to reduce out-of-bid situations and achieve low cost of using Cloud resources \cite{Mattess_Vecchiola_2010}. However, on the one hand, Cloud providers would increase spot prices to maximize profits if most consumers submit high bids \cite{Voorsluys_Buyya_2012}; on the other hand, the essence of incentive-compatible auction in the Cloud spot market would require consumers to bid their true valuations so as to obtain maximum utility \cite{Zhang_Gurses_2011}. Thus, the backend interactions between market participants could be too complicated for Cloud
consumers to understand psychologically \cite{Xu_Li_2013}, and correspondingly coming up with wise bids is clearly a nontrivial task for them. In fact, it has been identified that a main reason of the aforementioned consumers' hesitation is the complexity in determining suitable bids for obtaining spot resources \cite{Zaman_Grosu_2011}.

To address the challenge in employing Cloud spot services, we decided to focus on practical bidding techniques. Inspired by the Agile principle of gradual improvement, we started from investigating possible price predictions without taking into account consumer constraints or market competitions, and then developed a feedback control-based bidding mechanism. This paper introduces the whole work including both the mechanism development and its validation. In brief, we model the bidding system as an arccotangent function-based dynamics that accepts control signals as input and generates corresponding bids as output. Using a feedback loop, our bidding mechanism is supposed to imitate intuitive and fuzzy human activities to revise and issue new bids according to the previous errors. Compared with a set of other straightforward bidding strategies, this control-theoretical approach can achieve a relative balance between the success rate and the rationality of bidding. 

The contribution of this work is mainly threefold, as listed below.
\begin{itemize}
 \item{The bidding system modeling work introduces a lightweight mathematical model for dynamical systems within particular state spaces. Considering that the possible bids have a limited space between a price floor and a ceiling, the conventional state space models for this bidding system would be ordinary differential equations \cite{Astrom_Murray_2008}. Our study shows that an arccotangent function-based curve can match the real bidding situations well, and such a model could also be suitable for other similar systems.}

 \item{The feedback control-based mechanism suggests an intuitive approach to deal with the spot service bidding problem. Its essential idea is to imitate the fuzzy process of decision making in our normal lives. In addition to the convenience of implementation, this mechanism allows flexible adjustments at three different stages to help make more aggressive bidding (i.e.~generating lower-than-normal bids) or more conservative bidding (i.e.~generating higher-than-normal bids).}

 \item{Two generic features of bidding strategies (namely \textbf{success rate} and \textbf{relative rationality}) are defined and used to facilitate ``apple-to-apple" comparison between different options. To our best knowledge, this is the first study that considers these two features together to measure and compare bidding strategies. Given more consumer constraints and market conditions, they can be conveniently combined with other features like budget satisfaction for further trade-off comparison.}
\end{itemize}

The remainder of this paper is organized as follows. Section \ref{sec:relatedwork} summarizes the existing bidding strategies by roughly classifying them into three groups: white-box, grey-box, and black-box strategy types. Section \ref{sec:mechanism} elaborates our development details of this feedback control-based bidding mechanism. In addition to a simulation study, Section \ref{sec:validation} compares our bidding mechanism against five other straightforward bidding strategies with respect to their features success rate and relative rationality. Conclusions and some future work are discussed in Section \ref{sec:conclusion}.

\begin{figure*}[!t]
\centering
\includegraphics{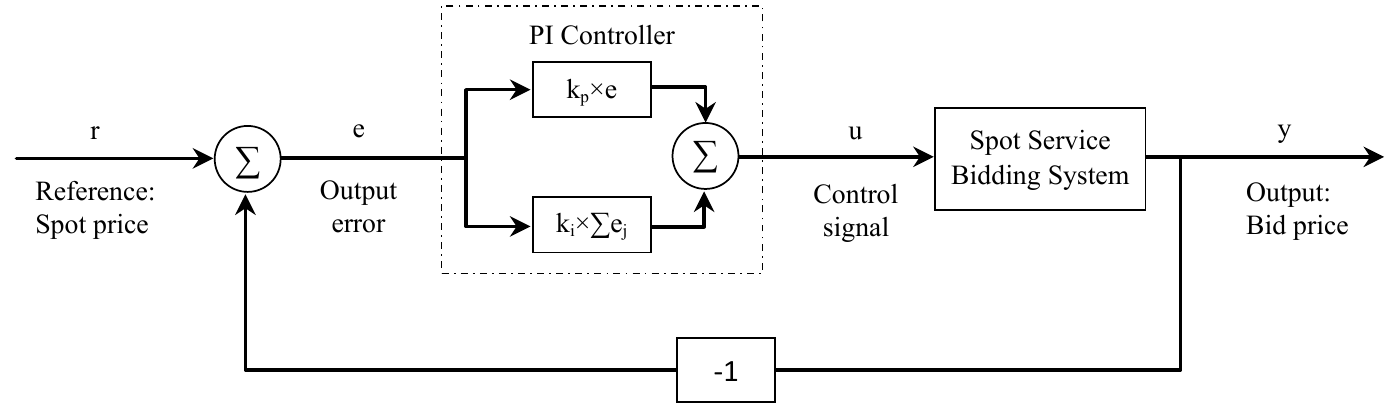}
\caption{Feedback control-based mechanism for bidding to use Cloud spot service. This mechanism takes historical spot prices as the reference signal $r$, uses a proportional-integral (PI) controller to transfer the output error $e$ into the control signal $u$, and eventually delivers bid prices as the output $y$.}
\label{fig_FeedbackModel}
\end{figure*}

\section{Rlated Work: A Classification of Bidding Strategies for Employing Cloud Spot Service}
\label{sec:relatedwork}
The existing relevant studies have investigated mainly three types of bidding strategies for employing Cloud spot service, and we name them as white-box, grey-box, and black-box strategies. This section explains these strategy types by giving typical examples respectively.

\subsection{White-box Bidding Strategies}

We consider a bidding strategy white-box when the strategy takes into account the interactions between different market participants. In other words, by including different participants' reactions, a white-box study would be concerned with the influence of bidding on the resulting spot prices. 

For example, Sowmya and Sundarraj \cite{Sowmya_Sundarraj_2012} focused on the competition between different bidders, and used a prisoner dilemma game to model the bidding scenario in the Cloud spot market. To facilitate analysis, they assumed only two bidders in the game, and proved that rational and self-interested consumers would converge on the Nash Equilibrium solution to procure spot resources. By analyzing Amazon's spot price data, the authors believed that the major bidders played the single shot classical prisoner dilemma game. Given the characteristics of Cloud spot market that might require consumers to bid in multiple rounds, the study proposed a co-operation bidding strategy that matches an iterated prisoner dilemma game. 

Zaman and Grosu \cite{Zaman_Grosu_2011} focused on the relationship between Cloud provider and consumers, and used combinatorial auctions to model the bidding activities in the Cloud spot market. They defined that a combinatorial auction mechanism was employed for Cloud providers to allocate spot resources. This mechanism sorts the bids in descending order of density of valuation, and greedily selects high-value bids to maximize the social welfare (i.e.~sum of valuation of all winning bidders). Since the resource allocation mechanism is supposed to be incentive-compatibility, the authors proposed a bidding strategy to generate truthful valuation of a bundle of spot resources within particular budgets, so as to satisfy Cloud consumers with the maximum utility. 

\subsection{Grey-box Bidding Strategies}
In contrast with white-box studies, grey-box strategies are proposed usually from an independent bidder's point of view, without considering the interactions with the other market participants. Instead, a typical focus of grey-box bidding is on a given set of constraints, e.g., workload, cost and/or availability criteria. 

For example, benefiting from a Price Transition Probability Matrix (PTPM) that characterizes different price transitions as time goes on, Tang et al.~\cite{Tang_Yuan_2012} developed a Constrained Markov Decision Process (CMDP) based bidding strategy to both minimize the monetary cost and satisfy the available time requirement. In particular, given the transition probability of each bid option in PTPM, the CMDP is used to model the computation of finding an optimal option with the predefined constraints, and the stochastic computation process is formulated as a linear programming and correspondingly solved in polynomial time. 

Based on a Markov model for spot price evolution, Zafer et al.~\cite{Zafer_Song_2012} defined spot resource employment as a cost minimization problem in the field of discrete-time stochastic Dynamic Programming, and utilized the relevant mathematical tools to achieve optimal bids from the consumer's perspective. Similarly, given a semi-Markovian process price model, Song et al.~\cite{Song_Zafer_2012} designed a profit-aware dynamic bidding (PADB) algorithm to make sequential bidding decisions for a job queue, and each decision only required the current job size and the current spot price. PADB is supposed to achieve a near-optimal bidding solution to the profit maximization problem from the service broker's perspective.

\subsection{Black-box Bidding Strategies}
Similar to the grey-box work, the black-box studies are not concerned with interactions between market participants either. In addition, unlike white-box bidding strategies, black-box bidding decisions are generally influenced by historical spot prices; and unlike grey-box strategies, black-box bidding does not take into account bidders' constraints. In essence, black-box bidding can be viewed as making straightforward predictions about future spot prices, without necessarily being aware of external conditions.

For example, Mazzucco and Dumas \cite{Mazzucco_Dumas_2011} treated spot price as a random variable under a normal distribution model, used the autocorrelation function (ACF) to measure the correlation of historical spot price with itself at different time points, and developed an ACF-based algorithm to realize spot price prediction for bidding. In an extreme case of theoretical discussion \cite{Voorsluys_Buyya_2012}, five simple bidding strategies were proposed and compared with each other, such as Minimum, Mean, High, Current, and On-demand. We particularly highlight these five strategies in Section \ref{subsec:comparison} for the purpose of validating our study.

In fact, a black-box bidding strategy could play a fundamental role in the other types of strategies. Considering that there is still a lack of a both convincing and practical bidding mechanism,  %For example, by using relative transition frequency to dynamically estimate the transition probability between two spot prices, the aforementioned PTPM \cite{Tang_Yuan_2012} for its corresponding grey-box bidding strategy is essentially an approach to spot price prediction. 
we start our investigation into bidding strategies from a fundamental study, as described in the following sections. Note that, as a black-box strategy at this current stage, our feedback control-based bidding mechanism does not include external constraints (or input signals, such as budget limit, workload priority, bidder competition, etc.).

\section{Feedback Control-based Bidding for Cloud Spot Service}
\label{sec:mechanism}
Our feedback control-based mechanism was initially designed by modeling spot service bidding as an input-output dynamical system. The system further employs a feedback loop to enable automatic adjustment of its input. The essential logic of the mechanism can be outlined into a block diagram, as illustrated in Fig.~\ref{fig_FeedbackModel}. In brief, this mechanism takes historical spot price as the reference signal $r$, uses a proportional-integral (PI) controller to transfer the output error $e$ into the control signal $u$, and eventually delivers bid price as the output $y$. Note that we are not concerned with any external disturbance in this case. The following subsections particularly explain our work on the system model and the PI controller followed by a simulation discussion.

\subsection{Modeling the Spot Service Bidding System}
\label{subsec:systemModeling}
Inspired by the input/output view of dynamics from electrical engineering \cite{Astrom_Murray_2008}, we focus on the input and output behaviors when modeling the bidding system: It is supposed to accept a control signal as input and output the corresponding bid. 

We start from considering the expected output. Firstly, Cloud consumers would expect successful bids. When consuming a Cloud spot service, as specified by Amazon \cite{Amazon_2015}, the requested spot resources can be granted only if the submitted bid exceeds the current spot price, otherwise the consumer has to resubmit new bids or wait for the request to be in-bid. Secondly, the generated bids should be rational. Suppose the activities in the Cloud spot market conform to the aforementioned incentive-compatible auction, the real spot price would reflect the true value of spot resources at a particular time; the accurate valuation of the spot service would then be equal to its price; and accordingly the rational bids should be close to the spot price. 

To achieve both successful and rational bids, we assume that the bidding system wants to use a control signal to correct its previous output errors. Intuitively, the system would decrease its bid in next round if the previous output is too high (for the purpose of rational bidding), while bidding higher if the previous bid is lower than the real spot price (for the purpose of successful bidding). In extreme cases, however, it is unacceptable to bid higher than the price of on-demand service, because Cloud consumers would rather directly employ on-demand resources if the spot service is too expensive; and it does not make sense to set a bid infinitely low either, because the Cloud provider could have a reserved price for the cost of developing and maintaining the spot service. In fact, Amazon's spot price traces have shown that there always exists a band in which the price fluctuation happens most of the time \cite{Mattess_Vecchiola_2010}.

Overall, given the intuitive discussion, it is possible to consider an arccotangent type of curve to satisfy the aforementioned constraints between control signal and bid price. On the other hand, it would be hard to fit those constraints with other functions like disrete steps or linear curves. Therefore, we formulate the bidding system into an arccotangent function-related model, as shown in Equation (\ref{eq_biddingSystemModel}).

\begin{equation}
\label{eq_biddingSystemModel}
  y=a+\frac{b-a}{\pi}\times \arccot(u)
\end{equation}

\noindent
where $u$ is the control signal, $y$ is the generated bid price, while $a$ and $b$ are the price floor and ceiling respectively for a particular Cloud spot service. By transferring the range from arccotangent's $\pi-0$ to $b-a$, this model never allows bids beyond the band from $a$ to $b$. 

\begin{figure}[!t]
\centering
\includegraphics[width=8.8cm]{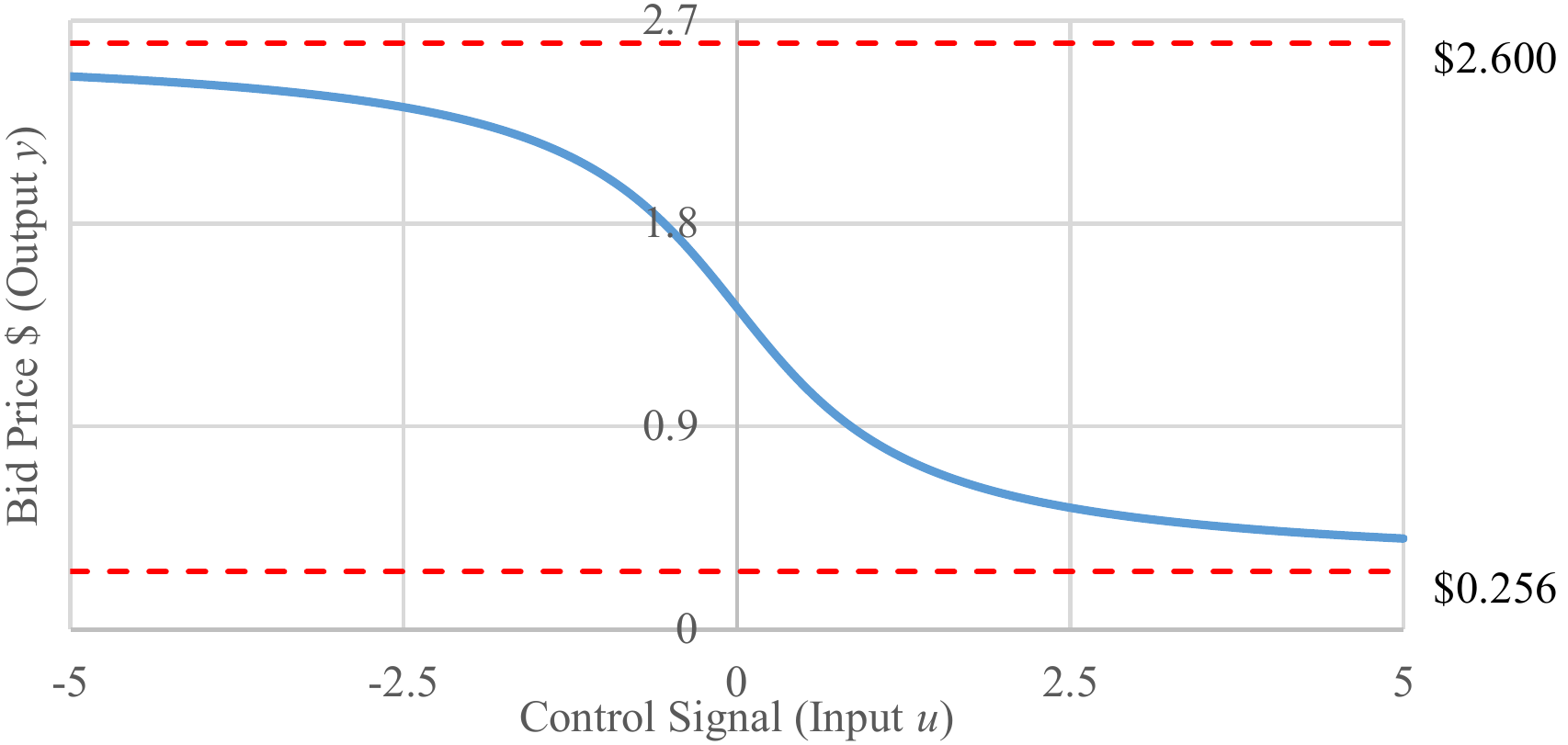}
\caption{Arccotangent function-based bidding system model. For Amazon's spot instance type g2.8xlarge, the price floor is \$0.256 and the price ceiling is \$2.600.}
\label{fig_ArccostShape}
\end{figure}

As such, the influence of control signals on bid prices can be plotted as illustrated in Fig.~\ref{fig_ArccostShape}. Here we take Amazon's EC2 spot instance type g2.8xlarge for example (cf.~Table \ref{tbl>spotInstance}), its lowest historical price is used as the price floor (i.e.~$a=0.256$), and the corresponding on-demand service price is defined as the price ceiling (i.e.~$b=2.600$).

\begin{table}[!t]\footnotesize
\renewcommand{\arraystretch}{1.3}
\centering
\caption{\label{tbl>spotInstance}Amazon's Instance Type G2.8xlarge}
\begin{tabular}{|l|c |c|c|c|}
\hline
\multirow{2}{*}{\textbf{Feature}} & \textbf{GPUs} & \textbf{vCPU} & \textbf{Mem (GiB)} & \textbf{Storage (GB)}\\
\cline{2-5}
& 4	& 32	& 60	& 2 x 120 SSD\\
\hline
\textbf{OS Usage} &\multicolumn{4}{l|}{Linux/UNIX usage}\\
\hline
\textbf{Region} &\multicolumn{4}{l|}{US East (N.~Virginia)}\\
\hline
\textbf{Hourly Price} &\multicolumn{4}{l|}{On-demand: \$2.600; Lowest Spot: \$0.256}\\
\hline
\end{tabular}
\end{table}

\subsection{Using a PI Controller to Generate Control Signal}
As mentioned previously, the control signal is supposed to help correct the past bidding errors. Naturally, we consider the control signal to be generated according to those errors. An error is defined as the difference between the output bid and the real spot price at a particular time, i.e., $e=r-y$ following the notations in Fig.~\ref{fig_FeedbackModel}. 

Firstly, we are concerned with the influence of the present error on the potential control signal. Recall that the bigger positive value of the error indicates the lower previous bid than it should be. Given the arccotangent-curve relationship between the control signal and the bid price (cf.~Fig.~\ref{fig_ArccostShape}), a bigger negative value of the control would then be expected in order to bid higher in the next round. Similarly, the bigger negative error would expect the bigger positive control for the lower bid next time. 

\begin{figure}[!t]
\centering
\includegraphics[width=8.8cm]{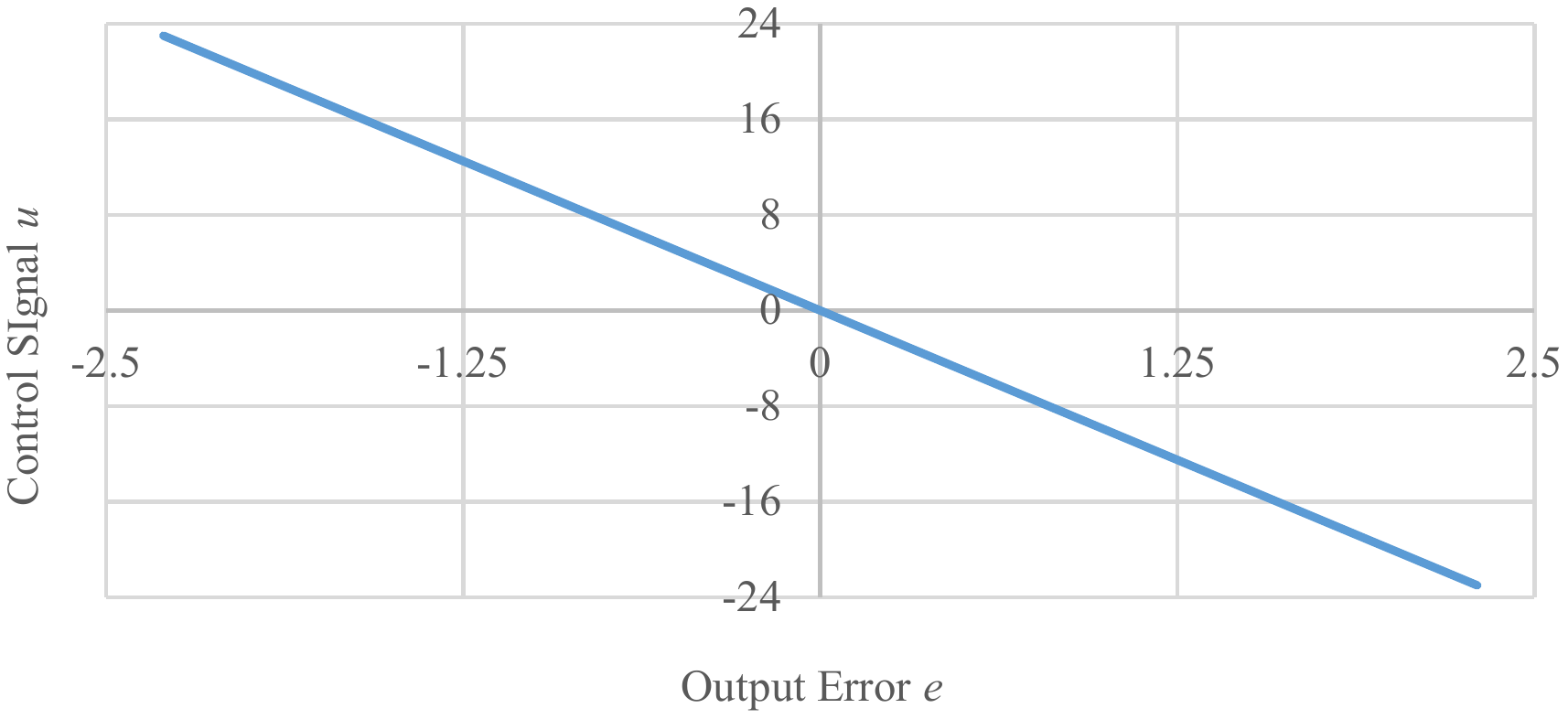}
\caption{An example proportional control with the proportional gain $-10$. Given the bidding for Amazon's spot instance type g2.8xlarge, the proportional band is $(-2.344,~2.344)$.}
\label{fig_ProportionalShape}
\end{figure}

Thus, it is clear that there is a negative correlation between the bid error and the control signal, and a proportional control can then fit in this case. We formulate this proportional control into Equation (\ref{eq_proportional}).

\begin{equation}
\label{eq_proportional}
  u_p=k_p\times e ~~~~ (k_p<0,~a-b<e<b-a)
\end{equation}

\noindent
where $k_p$ is the proportional gain for this control $u_p$, and its negative value guarantees the abovementioned error-control correlation. Moreover, considering that all the bids are in the interval $(a,~b)$ constrained by the system model (cf.~Equation (\ref{eq_biddingSystemModel})), the bid error $e$ has a proportional band $(a-b,~b-a)$ correspondingly. Following the example of the instance type g2.8xlarge, we show a typical proportional control with the gain $-10$ and the band $(-2.344,~2.344)$ in a linear graph in Fig.~\ref{fig_ProportionalShape}.

Secondly, we are concerned with the influence of the historical errors on the potential control signal. In fact, given the discussions in the system modeling work (cf.~Section \ref{subsec:systemModeling}), the proportional control essentially issues intuitive trials, because it is impossible to predetermine to what extent we can match a new spot price by fixing the previous error. Consequently, we would have to track the consecutive past errors to observe the effects of the existing proportional control trials. For example, the gradually increasing value of the integral of errors may indicate that the historical trials have been too conservative. In other words, the historical errors all together could comprise further information to help improve new bidding for Cloud spot service.

\begin{figure}[!t]
\centering
\includegraphics[width=8.8cm]{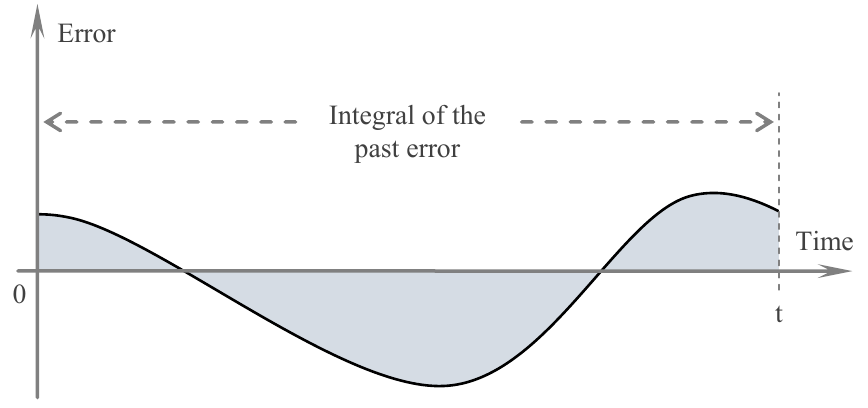}
\caption{The shaded portion is the integral of bid errors up to time $t$, which indicates the historical information that can be used for the integral component $u_i$ of the overall control $u$.}
\label{fig_IntegralShape}
\end{figure}

Therefore, we decided to employ an integral control to further utilize the historical information, as visualized in the shaded portion in Fig.~\ref{fig_IntegralShape}. Following the similar discussion for the proportional control, the integral control action would also be proportional to the integral of errors, as formulated in Equation (\ref{eq_integral}). 

%\begin{empheq}{align}
%\begin{eqnarray}
\begin{equation}
\label{eq_integral}
\begin{aligned}
   &u_i(t)=k_i\times \int_0^t e(\tau)\,d\tau \approx k_i\times \sum_{j=0}^t{e_j} \\ 
	&(k_i<0,~a-b<e_j<b-a)
\end{aligned}
\end{equation}
%\end{empheq}

\noindent
where $k_i$ is the integral gain for the control $u_i$, and its negative value also implies the negative correlation between the control signal and the integral of errors. Considering that Amazon tends to hold a period of time between different price points \cite{Wee_2011}, we replace $\displaystyle\int_0^t e(\tau)\,d\tau$ with $\sum\limits_{j=0}^t{e_j}$ to calculate the integral of discrete-time errors up to time $t$, and clearly each historical error $e_j$ must have been in the aforementioned interval $(a-b,~b-a)$.

\begin{figure*}[!t]
\centering
\includegraphics[width=18cm]{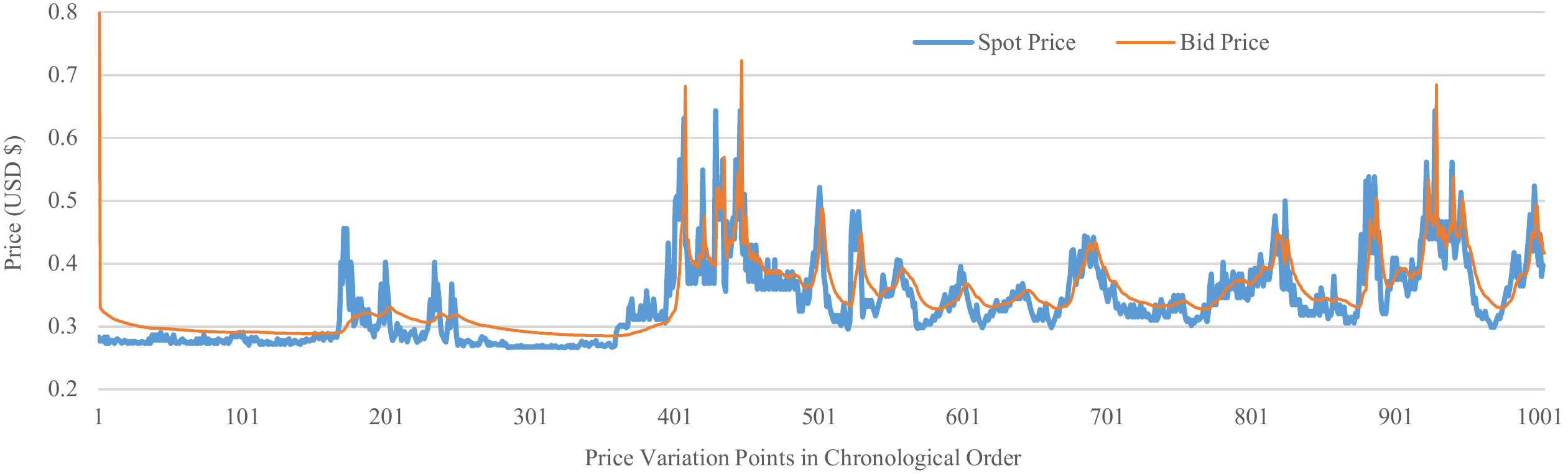}
\caption{Simulation of the feedback control-based bidding mechanism with respect to Amazon's spot price variation trace between 2015-05-03 00:20:06 and 2015-05-14 23:36:44 (instance type: g2.8xlarge, OS type: Linux/UNIX, zone: us-east-1b).}
\label{fig_Simulation4FBC}
\end{figure*}

Overall, based on these two types of control, a PI controller with the input-output relation between bid error and control signal can be defined as:
~~\\
\begin{equation}
\label{eq_PIcontroller}
   u=u_p+u_i \approx k_p\times e + k_i\times \sum_{j=0}^t{e_j} 
\end{equation}

The summary control action is thus composed of the proportional and integral feedbacks. In particular, the proportional feedback can be viewed as the fine tuning to adjust bidding based on the present error, while the integral feedback can be viewed as the coarse tuning to adjust bidding based on the accumulated past errors. 

\subsection{Simulation of the Feedback Control-based Bidding Mechanism}
In addition to the previous qualitative discussion about designing the feedback control-based bidding mechanism, we now use a simulation to quantitatively demonstrate how this mechanism works. To facilitate the simulation, we implement the bidding mechanism into executable codes, as specified in Algorithm \ref{BiddingAlgorithm}. This straightforward algorithm also shows that the deployment of our bidding mechanism would not be difficult. Note that although Algorithm \ref{BiddingAlgorithm} is for simulation, the last element in the output array of bid prices can be directly used for the new round of bidding at time $t+1$. 

When it comes to the input parameters, we take into account the instance type g2.8xlarge that is recently available in the Cloud spot market. In detail, we set the price floor $a$ and ceiling $b$ as 0.256 and 2.600 respectively (cf.~Fig.~\ref{fig_ArccostShape}); for the conciseness of this simulation, we assign 10 to both the proportional gain $k_p$ and the integral gain $k_i$ without sophisticated parameter tuning; and the half price of the corresponding on-demand service is used for the initial random bidding.

As for the historical spot prices, we use Amazon's Command Line Interface (CLI) tool ec2-describe-spot-price-history \cite{Amazon_2015_tool} to collect spot price traces of the instance type g2.8xlarge. Without loss of generality, we select 1001 consecutive spot price records (with the timestamp between 2015-05-03 and 2015-05-14) as the input array for this simulation.\footnote{The complete spot price trace with 3849 records of the spot instance type g2.8xlarge has been shared online: \url{https://docs.google.com/spreadsheets/d/18iL2qYCpqsYyx-2hgzp950Fi8nP7K8xQhse_0gP8xjU/}}

The simulation result of our feedback control-based bidding is plotted together with the corresponding spot price trace in Fig.~\ref{fig_Simulation4FBC}. Recall that this bidding mechanism essentially imitates human activities to issue intuitive trials according to the previous errors. Since the bidding errors would never disappear due to spot price variations, the intuitive trials should also reflect the changes in spot prices. Driven by the feedback control signals, the simulation shows that the bid trajectory roughly follows the spot price trace. It is then reasonable to assume that the bidding mechanism works, at least to some intuitive extent. To better validate this feedback control-based bidding mechanism, we compare it against a set of other bidding strategies, as specified in the following section.

\begin{algorithm}[!t]\footnotesize
%\algsetup{linenosize=\small}
\caption{Feedback Control-based Bidding Mechanism}
\label{BiddingAlgorithm}
        \textbf{Input:} Array of historical spot prices $P=(p_1, p_2, p_3,...,p_t)$, price floor $a$, price ceiling $b$, proportional gain $k_p$, integral gain $k_i$, random bid price $\textit{bid}_1$. \\
        \textbf{Output:} Array of bid prices $\textit{BID}=(\textit{bid}_2, \textit{bid}_3,...,\textit{bid}_t,\textit{bid}_{t+1})$. 
\begin{algorithmic}[1]
\Procedure{Initialization}{}
\State Present output error $e \gets 0$
\State Summary output error $e_\text{\textit{sum}} \gets 0$
\State Control signal $u \gets 0$
\State Bid for next round $\textit{bid} \gets \textit{bid}_1$
\State Array of bid prices $\textit{BID} \gets \emptyset$
\EndProcedure
\Procedure{Bid Generation Loop}{}
\For{$j=1,2,3,...,t$ }
\State $e \gets p_j-\textit{bid}$
\State $e_\text{\textit{sum}} \gets e_\text{\textit{sum}} + e$
\State $u \gets k_p \times e + k_i \times e_\text{\textit{sum}}$
\State $\textit{bid} \gets a+(b-a)\times \arccot(u)/\pi$
\State $\textit{BID} \gets \textit{BID} \cup \textit{bid}$
\EndFor
\EndProcedure
\State \Return \textit{BID}
\end{algorithmic}
\end{algorithm}

\section{Validation of the Feedback Control-based Bidding Mechanism}
\label{sec:validation}
Instead of implementing practical applications, we validate our bidding mechanism by contrasting it with other comparable bidding strategies. Note that, to be compatible with the descriptions in the existing studies (e.g., \cite{Voorsluys_Buyya_2012}), we also treat ``bidding mechanism" and ``bidding strategy" as interchangeable terms in this paper. 

\subsection{Two Generic Features of a Bidding Mechanism}
To make ``apple-to-apple" comparison, following the discussion in Section \ref{subsec:systemModeling}, we focus on two features of a particular bidding mechanism, namely \textbf{success rate} and \textbf{relative rationality}. 

The \textbf{success rate} refers to the percentage of in-bid events among all the bid trials during a period of time. An in-bid event indicates that the bid price is greater than or equal to the corresponding spot price at a particular time. In fact, in the Cloud spot market, a consumer's request can be satisfied if and only if the consumer's bid price exceeds the current spot price. Since it is natural for Cloud consumers to expect successful bids, we claim that a better bidding strategy should have higher success rate than another. The calculation of success rate is formulated into Equation (\ref{eq_successRate}).

\begin{equation}
\label{eq_successRate}
   sr=\frac{\sum\limits_{i=1}^t(i^{th}~bid~price \geq i^{th}~spot~price~?~1:0)}{t} \times 100\%
\end{equation} 

\noindent
where $sr$ is the abbreviation for ``success rate", and $t$ represents the times of spot price variations. Note that for the purpose of validation we suppose each spot price variation triggers a bid trial for the next round of bidding, without considering the fact that the successfully requested spot resources can be allocated until either the consumer intentionally terminates the service usage or the spot price increases above the previous bid price.

As for the rationality, recall that the economics principles would constrain rational bids to be close to the spot price (cf.~Section \ref{subsec:systemModeling}). We naturally consider that the rationality of a bidding strategy is inversely proportional to the distance between the generated bid trajectory and its related spot price trace. Given the discrete-time variations of spot price, a simple way to calculate the aforementioned distance can be defined as shown in Equation (\ref{eq_distance}). 

\begin{equation}
\label{eq_distance}
   d=\sum_{i=1}^t\left|i^{th}~bid~price - i^{th}~spot~price\right| 
\end{equation} 

Correspondingly, the rationality can simply be defined as $r=1/d$ for instance. An assumption here is that the ideal bidding strategy with zero distance ($d=0$) does not exist in practice. Then, as discussed previously, we claim that a better bidding strategy should have higher rationality than another.

To facilitate comparison within an appropriate scale, we further come up with using \textbf{relative rationality} to measure a group of bidding strategies, as expressed in Equation (\ref{eq_rationality}).

\begin{equation}
\label{eq_rationality}
   rr_j=\frac{\min\limits_{d_i\in D}\{d_i\}}{d_j} ~~~ (i,j=1,2,...,n)
\end{equation} 

\noindent
where $rr_j$ refers to the relative rationality of the $j^{th}$ bidding strategy among $n$ ones, $d_j$ is the distance between the $j^{th}$ bid trajectory and the same spot price trace, and $D$ indicates the distance set with respect to those $n$ bidding strategies. By transforming rationality into relative rationality, the bidding strategy with the shortest distance obtains the score $1$, while the others have scores over the interval $(0,~1)$. It is notable that the comparison condition is still the same, i.e.~a better bidding strategy should have higher relative rationality than another.

\begin{figure*}[!t]
\centering
\includegraphics[width=18cm]{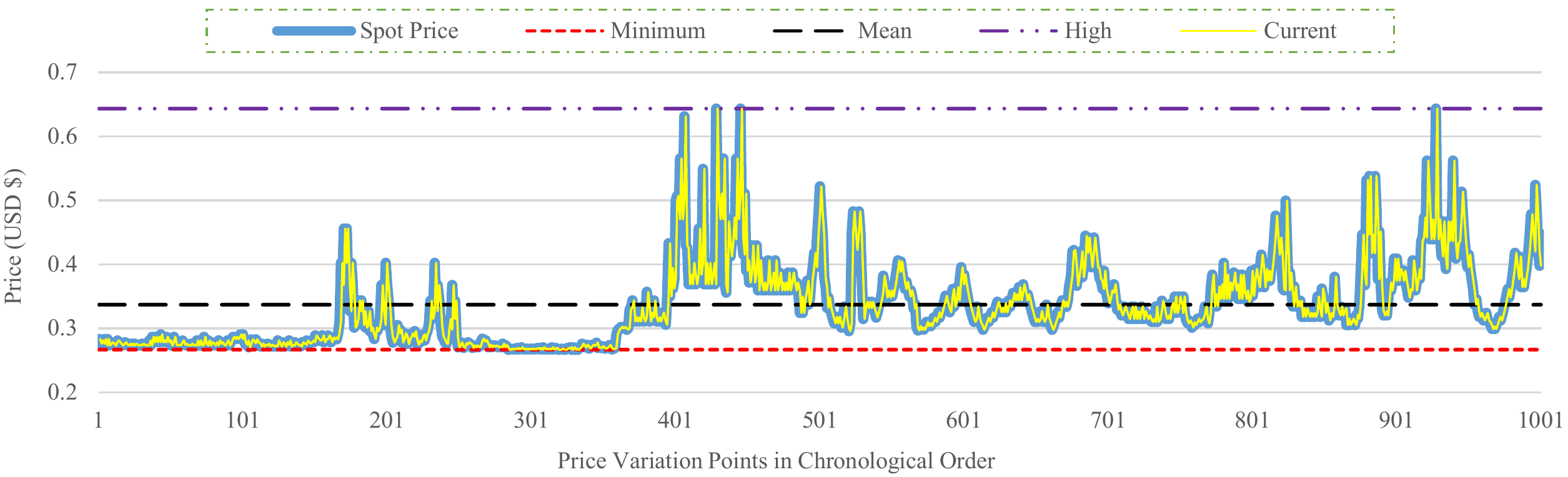}
\caption{Simulation of the four black-box bidding strategies with respect to Amazon's spot price variation trace between 2015-05-03 00:20:06 and 2015-05-14 23:36:44 (instance type: g2.8xlarge, OS type: Linux/UNIX, zone: us-east-1b).}
\label{fig_Simulation4Others}
\end{figure*}

\subsection{Comparison between Different Bidding Mechanisms}
\label{subsec:comparison}
As introduced in Section \ref{sec:relatedwork}, five black-box bidding strategies have been proposed in the existing study \cite{Voorsluys_Buyya_2012}. Here we briefly rephrase them as follows.

\begin{itemize}
 \item{\textit{Minimum:} The bid price is set as the minimum value observed in the spot price history.}
 \item{\textit{Mean:} The bid price is set as the mean of all values in the spot price history.}
 \item{\textit{High (Maximum):} The bid price is set as a value higher than any price observed. To distinguish from the \textit{On-demand} strategy, we define the \textit{high value} as the maximum value observed in the spot price history.}
 \item{\textit{Current:} The bid price is set as the value of the current spot price.}
 \item{\textit{On-demand:} The bid price is set as the value of the corresponding on-demand service price.}
\end{itemize}

By applying these bidding strategies to the same spot price trace, we simulate their bidding activities and generate bid trajectories, as illustrated in Fig.~\ref{fig_Simulation4Others}. Note that the \textit{On-demand} bidding result is invisible in this figure due to its far location from the others.

\begin{figure}[!t]
\centering
\includegraphics[width=8.5cm]{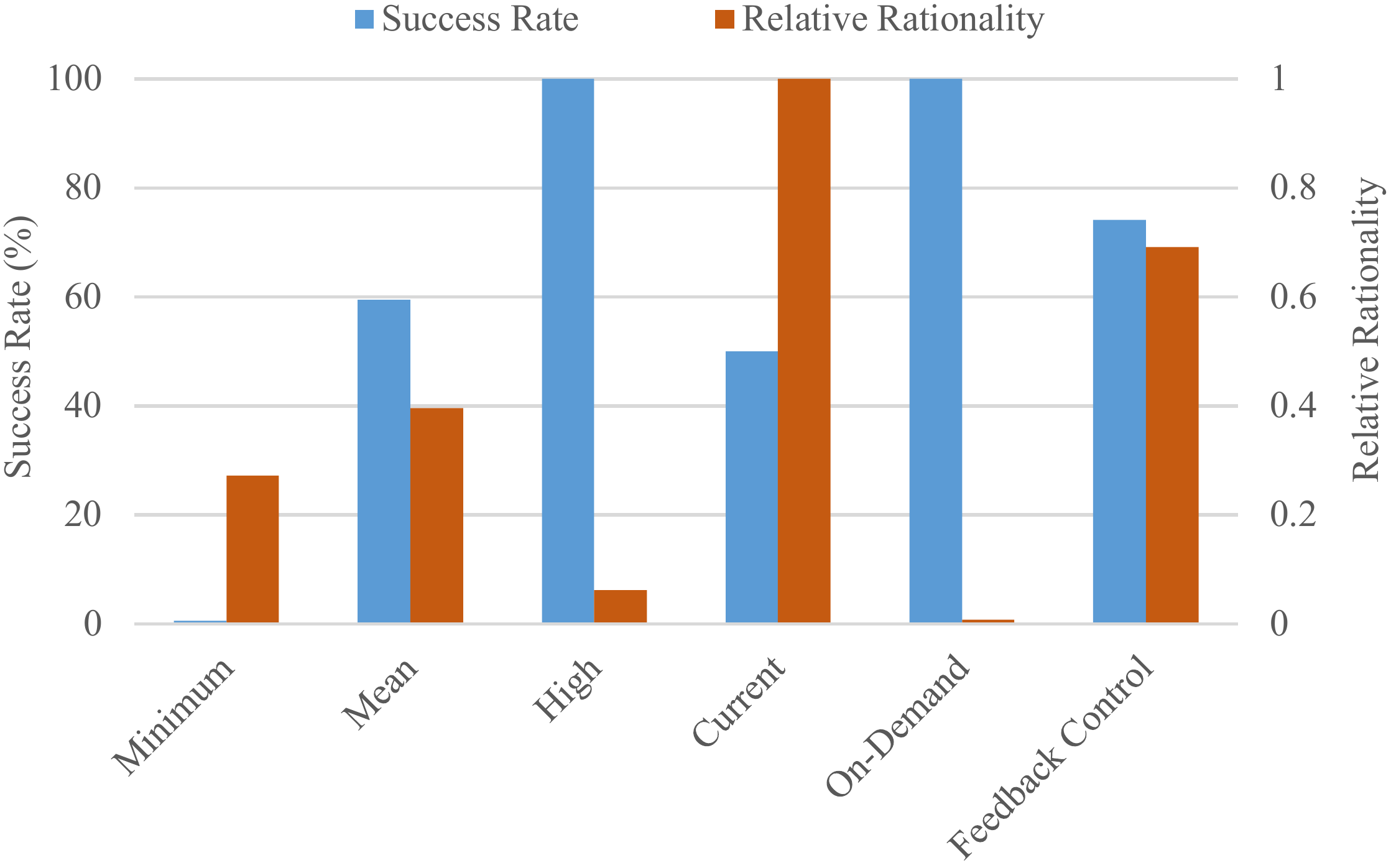}
\caption{Comparison between different bidding strategies in terms of Success Rate and Relative Rationality. The calculation is based on Amazon's spot price variation trace between 2015-05-03 00:20:06 and 2015-05-14 23:36:44 (instance type: g2.8xlarge, OS type: Linux/UNIX, zone: us-east-1b).}
\label{fig_Validation}
\end{figure}

Given the generated bid trajectories over the same time span, we respectively calculate the success rate and relative rationality for the five bidding strategies together with our bidding mechanism respectively. The comparison result is shown in Fig.~\ref{fig_Validation}. In detail, if employing the \textit{Minimum} strategy to try to take advantage of the cheapest spot service, the success rate of bidding could be extremely low. On the contrary, if employing the \textit{High} or even \textit{On-demand} strategies to pursue $100\%$ success rate, the bidding rationality would be unacceptably low. In fact, as modeled by using the Prisoner Dilemma game \cite{Sowmya_Sundarraj_2012}, no one can always win the bidding if every consumer tries to bid high. On the other hand, although the \textit{Mean} and \textit{Current} seem to be more reasonable bidding strategies, both of them still suffer from the relatively significant imbalance between success rate and rationality. In contrast, the feedback control essentially delivers a trade-off mechanism that gives consideration to both bidding rationality and success rate.

More importantly, our feedback control-based bidding mechanism is more flexible than the existing fixed bidding strategies. As a matter of fact, without significantly violating rationality, Cloud consumers can adjust their bidding at three different stages when employing this mechanism. 

\begin{itemize}
 \item{\textit{Pre-adjustment:} The bidding can be adjusted by modifying reference signals (i.e.~changing the historical spot prices). For example, a consumer can increase the reference signal by adding two more cents to the historical spot prices, so as to make conservative bidding to enhance the success rate; or the consumer may use aggressive bidding to enjoy cheaper spot service by decreasing reference signals.}

 \item{\textit{Control Adjustment:} The bidding can be adjusted by modifying the controller gains. For example, a consumer can increase the proportional and integral gains (i.e.~$k_p$ and $k_i$) to improve the PI controller's sensitivity about present and historical errors respectively, or vice versa.}

 \item{\textit{Post-adjustment:}  The bidding can be adjusted by directly modifying the output bids. It is clear that the post-adjustment is also suitable for the other bidding strategies. For example, a consumer can give two more cents top-up to the generated bid to make conservative bidding, or reducing the suggested bid price to conduct aggressive bidding.}
\end{itemize}

Considering the combination explosion of available parameters, we do not illustrate the detailed effects of different adjustments.

\section{Conclusions and Future Work}
\label{sec:conclusion}
The current Cloud providers largely use three types of pricing schemes to sell their on-demand, reserved and spot services respectively. In particular, spot pricing has been claimed to be the most cost-effective scheme among all the options for Cloud consumers. Unfortunately, the existing Cloud spot services seem not to be popular yet in practice, and one of the main reasons has been identified to be the complexity and difficulty in determining suitable bids. Therefore, we focused on practical bidding strategies and developed a feedback control-based mechanism to facilitate bidding for Cloud spot service.

By using Amazon's historical spot price trace to perform a set of simulations, we show that our mechanism obtains a better trade-off between bidding rationality and success rate than the other five comparable strategies. Moreover, since the feedback control here has been designed to conform to the bidder's natural and fuzzy intuition, this bidding mechanism would be easily comprehensible for Cloud consumers. Benefiting from a straightforward algorithm, the simulation also indicates that this mechanism is deployment friendly. 

In addition to the bidding mechanism itself, this whole work has enlightened us about the possibility and feasibility of using control-theoretical approaches to deal with different problems in the Cloud spot market. For example, Cloud spot pricing could also be modeled as a control problem from the provider's perspective, because the spot market has granted Cloud providers the power of intentionally terminating spot services through controlling spot prices. In other words, we suggest using this study as an inspiration to reveal further research opportunities.

As for the near future, our potential work will be unfolded along two directions. The first, we plan to refine this feedback control-based bidding mechanism by tuning the system parameters and improving the controller. In particular, the controller could be improved by optimizing its sensitivity about bid errors, and by changing its PI control into a proportional-integral-derivative (PID) one. The second, we plan to gradually upgrade our bidding mechanism by taking into account consumer constraints and market competitions. Recall that black-box bidding strategies can not only work alone, but also play a fundamental role in other types of strategies. It would then be natural and logical to extend this work into more sophisticated bidding mechanisms to cover more external conditions.


\section*{Acknowledgment}

This work is supported by the Swedish Research Council
(VR) for the project ``Cloud Control", and through the
LCCC Linnaeus and ELLIIT Excellence Centers.

\bibliographystyle{IEEEtran}
%\setbiblabelwidth{99}
\bibliography{CloudCom_Ref}

\begin{thebibliography}{10}
\providecommand{\url}[1]{#1}
\csname url@samestyle\endcsname
\providecommand{\newblock}{\relax}
\providecommand{\bibinfo}[2]{#2}
\providecommand{\BIBentrySTDinterwordspacing}{\spaceskip=0pt\relax}
\providecommand{\BIBentryALTinterwordstretchfactor}{4}
\providecommand{\BIBentryALTinterwordspacing}{\spaceskip=\fontdimen2\font plus
\BIBentryALTinterwordstretchfactor\fontdimen3\font minus
  \fontdimen4\font\relax}
\providecommand{\BIBforeignlanguage}[2]{{%
\expandafter\ifx\csname l@#1\endcsname\relax
\typeout{** WARNING: IEEEtran.bst: No hyphenation pattern has been}%
\typeout{** loaded for the language `#1'. Using the pattern for}%
\typeout{** the default language instead.}%
\else
\language=\csname l@#1\endcsname
\fi
#2}}
\providecommand{\BIBdecl}{\relax}
\BIBdecl

\bibitem{Weinhardt_Anandasivam_2009}
C.~Weinhardt, A.~Anandasivam, B.~Blau, N.~Borissov, T.~Meinl, W.~Michalk, and
  J.~St{\"o}{\ss}er, ``Cloud computing – a classification, business models, and
  research directions,'' \emph{Business and Information Systems Engineering},
  vol.~1, no.~5, pp. 391--399, October 2009.

\bibitem{Al-Roomi_2013}
M.~Al-Roomi, S.~Al-Ebrahim, S.~Buqrais, and I.~Ahmad, ``Cloud computing pricing
  models: {A} survey,'' \emph{International Journal of Grid and Distributed
  Computing}, vol.~6, no.~5, pp. 93--106, 2013.

\bibitem{Xu_Li_2013}
H.~Xu and B.~Li, ``Dynamic {Cloud} pricing for revenue maximization,''
  \emph{IEEE Transactions on Cloud Computing}, vol.~1, no.~2, pp. 158--171,
  July-December 2013.

\bibitem{Wang_Qi_2013}
P.~Wang, Y.~Qi, D.~Hui, L.~Rao, and X.~Lin, ``Present or future: {Optimal}
  pricing for spot instances,'' in \emph{Proceedings of the 33rd International
  Conference on Distributed Computing Systems (ICDCS 2013)}.\hskip 1em plus
  0.5em minus 0.4em\relax Philadelphia, USA: IEEE Computer Society, 8-11 July
  2013, pp. 410--419.

\bibitem{Song_Yao_2013}
K.~Song, Y.~Yao, and L.~Golubchik, ``Exploring the profit-reliability trade-off
  in {Amazon's} spot instance market: {A} better pricing mechanism,'' in
  \emph{Proceedings of the 21st IEEE/ACM International Symposium on Quality of
  Service (IWQoS 2013)}.\hskip 1em plus 0.5em minus 0.4em\relax Montreal,
  Canada: IEEE Press, 3-4 June 2013, pp. 1--10.

\bibitem{Ben-Yehuda_2013}
O.~A. Ben-Yehuda, M.~Ben-Yehuda, A.~Schuster, and D.~Tsafrir, ``Deconstructing
  {Amazon EC2} spot instance pricing,'' \emph{ACM Transactions on Economics and
  Computation}, vol.~1, no.~3, pp. 1--20, September 2013.

\bibitem{Javadi_Thulasiram_2013}
B.~Javadi, R.~K. Thulasiram, and R.~Buyya, ``Characterizing spot price dynamics
  in public {Cloud} environments,'' \emph{Future Generation Computer Systems},
  vol.~29, no.~4, pp. 988--999, June 2013.

\bibitem{Leslie_Lee_2013}
L.~M. Leslie, Y.~C. Lee, P.~Lu, and A.~Y. Zomaya, ``Exploiting performance and
  cost diversity in the {Cloud},'' in \emph{Proceedings of the 6th IEEE
  International Conference Cloud Computing (CLOUD 2013)}.\hskip 1em plus 0.5em
  minus 0.4em\relax Santa Clara, CA, USA: IEEE Computer Society, 28 June - 3
  July 2013, pp. 107--114.

\bibitem{Mattess_Vecchiola_2010}
M.~Mattess, C.~Vecchiola, and R.~Buyya, ``Managing peak loads by leasing
  {Cloud} infrastructure services from a spot market,'' in \emph{Proceedings of
  the 12th IEEE International Conference on High Performance Computing and
  Communications (HPCC 2010)}.\hskip 1em plus 0.5em minus 0.4em\relax
  Melbourne, VIC, Australia: IEEE Computer Society, 1-3 September 2010, pp.
  180--188.

\bibitem{Chohan_Castillo_2010}
N.~Chohan, C.~Castillo, M.~Spreitzer, M.~Steinder, A.~Tantawi, and C.~Krintz,
  ``See spot run: {Using} spot instances for {MapReduce} workflows,'' in
  \emph{Proceedings of the 2nd USENIX conference on Hot topics in cloud
  computing (HotCloud 2010)}.\hskip 1em plus 0.5em minus 0.4em\relax Boston,
  MA, USA: USENIX Association, 22 June 2010, pp. 1--7.

\bibitem{Ostermann_Prodan_2012}
S.~Ostermann and R.~Prodan, ``Impact of variable priced {Cloud} resources on
  scientific workflow scheduling,'' in \emph{Proceedings of the 18th
  International Conference on Parallel Processing (Euro-Par 2012)}.\hskip 1em
  plus 0.5em minus 0.4em\relax Rhodes Island, Greece: Springer-Verlag, 27-31
  August 2012, pp. 350--362.

\bibitem{Voorsluys_Buyya_2012}
W.~Voorsluys and R.~Buyya, ``Reliable provisioning of spot instances for
  compute-intensive applications,'' in \emph{Proceedings of the 26th IEEE
  International Conference on Advanced Information Networking and Applications
  (AINA 2012)}.\hskip 1em plus 0.5em minus 0.4em\relax Fukuoka, Japan: IEEE
  Computer Society, 26-29 March 2012, pp. 542--549.

\bibitem{Zhang_Gurses_2011}
Q.~Zhang, E.~G{\"u}rses, R.~Boutaba, and J.~Xiao, ``Dynamic resource allocation
  for spot markets in {Clouds},'' in \emph{Proceedings of the 11th USENIX
  Workshop on Hot Topics in Management of Internet, Cloud, and Enterprise
  Networks and Services (Hot-ICE 2011)}.\hskip 1em plus 0.5em minus 0.4em\relax
  Boston, MA, USA: USENIX Association, 29 March 2011, pp. 1--6.

\bibitem{Zaman_Grosu_2011}
S.~Zaman and D.~Grosu, ``Efficient bidding for virtual machine instances in
  {Clouds},'' in \emph{Proceedings of the 4th IEEE International Conference on
  Cloud~Computing (CLOUD 2011)}.\hskip 1em plus 0.5em minus 0.4em\relax
  Washington, DC, USA: IEEE Computer Society, 4-9 July 2011, pp. 41--48.

\bibitem{Astrom_Murray_2008}
K.~J. {\AA}str{\"o}m and R.~M. Murray, \emph{Feedback Systems: An Introduction
  for Scientists and Engineers}.\hskip 1em plus 0.5em minus 0.4em\relax
  Princeton, New Jersey: Princeton University Press, April 2008.

\bibitem{Sowmya_Sundarraj_2012}
K.~Sowmya and R.~P. Sundarraj, ``Strategic bidding for {Cloud} resources under
  dynamic pricing schemes,'' in \emph{Proceedings of the 2012 International
  Symposium on Cloud and Services Computing (ISCOS 2012)}.\hskip 1em plus 0.5em
  minus 0.4em\relax Mangalore, India: IEEE Computer Society, 17-18 December
  2012, pp. 25--30.

\bibitem{Tang_Yuan_2012}
S.~Tang, J.~Yuan, and X.-Y. Li, ``Towards optimal bidding strategy for {Amazon
  EC2 Cloud} spot instances,'' in \emph{Proceedings of the 5th IEEE
  International Conference on Cloud Computing (CLOUD 2012)}.\hskip 1em plus
  0.5em minus 0.4em\relax Honolulu, Hawaii, USA: IEEE Computer Society, 24-29
  June 2012, pp. 91--98.

\bibitem{Zafer_Song_2012}
M.~Zafer, Y.~Song, and K.-W. Lee, ``Optimal bids for spot {VMs} in a {Cloud}
  for deadline constrained jobs,'' in \emph{Proceedings of the 5th IEEE
  International Conference on Cloud Computing (CLOUD 2012)}.\hskip 1em plus
  0.5em minus 0.4em\relax Honolulu, Hawaii, USA: IEEE Computer Society, 24-29
  June 2012, pp. 75--82.

\bibitem{Song_Zafer_2012}
Y.~Song, M.~Zafer, and K.-W. Lee, ``Optimal bidding in spot instance market,''
  in \emph{Proceedings of the 31st Annual IEEE International Conference on
  Computer Communications (INFOCOM 2012)}.\hskip 1em plus 0.5em minus
  0.4em\relax Orlando, Florida, USA: IEEE Press, 25-30 March 2012, pp.
  190--198.

\bibitem{Mazzucco_Dumas_2011}
M.~Mazzucco and M.~Dumas, ``Achieving performance and availability guarantees
  with spot instances,'' in \emph{Proceedings of the 13th IEEE International
  Conference on High Performance Computing and Communications (HPCC
  2011)}.\hskip 1em plus 0.5em minus 0.4em\relax Banff, Canada: IEEE Computer
  Society, 2-4 September 2011, pp. 296--303.

\bibitem{Amazon_2015}
Amazon, ``Amazon {EC2} spot instances,''
  \url{https://aws.amazon.com/ec2/purchasing-options/spot-instances/}, March
  2015.

\bibitem{Wee_2011}
S.~Wee, ``Debunking real-time pricing in {Cloud} computing,'' in
  \emph{Proceedings of the 11th IEEE/ACM International Symposium on Cluster,
  Cloud and Grid Computing (CCGrid 2011)}.\hskip 1em plus 0.5em minus
  0.4em\relax Newport Beach, CA, USA: IEEE Computer Society, 23-26 May 2011,
  pp. 585--590.

\bibitem{Amazon_2015_tool}
Amazon, ``ec2-describe-spot-price-history,''
  \url{http://docs.aws.amazon.com/AWSEC2/latest/CommandLineReference/ApiReference-cmd-DescribeSpotPriceHistory.html},
  March 2015.

\end{thebibliography}

\end{document}